\documentstyle[aas2pp4,psfig]{article} 
\slugcomment{Accepted for publication in the Astrophysical Journal}
\begin{document} 

\title{The Outbursts and Orbit of the Accreting Pulsar GS 1843-02 = 2S 1845-024}

\author{Mark H. Finger\altaffilmark{1,2}, 
Lars Bildsten\altaffilmark{3}, 
Deepto Chakrabarty\altaffilmark{4},  
Thomas A. Prince\altaffilmark{5},  
D. Matthew Scott\altaffilmark{1,2}, 
Colleen A. Wilson\altaffilmark{1},  
Robert B. Wilson\altaffilmark{1} and 
S. Nan Zhang\altaffilmark{1,6}
}

\altaffiltext{1}{Space Science Laboratory, NASA/Marshall Space Flight Center,
ES 84,  Huntsville, AL 35812; mark.finger@msfc.nasa.gov, 
scott@gibson.msfc.nasa.gov, colleen.wilson@msfc.nasa.gov,
robert.b.wilson@msfc.nasa.gov, shuang.zhang@msfc.nasa.gov}

\altaffiltext{2}{Universities Space Research Association, 4950 Corporate Dr.,
Suite 100, Huntsville, AL 35806}

\altaffiltext{3}{Departments of Physics and Astronomy, 366 LeConte Hall,
University of California, Berkeley, CA 94720; bildsten@fire.berkeley.edu}

\altaffiltext{4}{Department of Physics and Center for Space Research, 
Massachusetts Institute of Technology, Cambridge MA 02139; deepto@space.mit.edu}

\altaffiltext{5}{Space Radiation Laboratory, California Institute of Technology,
Pasadena, CA 91125; prince@srl.caltech.edu}

\altaffiltext{6}{Department of Physics, University of Alabama in Huntsville,
Huntsville, AL 35899}
\begin{abstract} 
 
We present observations of a series of 10 outbursts of pulsed hard
X-ray flux from the transient 10.6 mHz accreting pulsar GS 1843-02, 
using the Burst and Transient Source Experiment on the 
Compton Gamma Ray Observatory. These outbursts
occurred regularly every 242 days, coincident with the ephemeris of the
periodic transient GRO J1849-03 (\cite{Zhang96})), which has recently been
identified with the SAS 3 source 2S 1845-024 (\cite{Soffitta98}).  Our pulsed
detection provides the first clear identification of  GS 1843-02 with 2S
1845-024.  We present a pulse timing analysis which shows that the 2S 1845-024
outbursts occur near the periastron passage of the neutron star's  highly
eccentric ($e = 0.88\pm0.01$)  $242.18\pm0.01$ day period binary orbit about a
high mass ($M_{\rm c} > 7 M_\odot$) companion. The orbit and transient outburst
pattern strongly suggest the pulsar is in a binary system with a Be star. 
Our observations show a long-term spin-up trend, with
most of the spin-up occurring during the outbursts. From the measured spin-up
rates and inferred luminosities we conclude that an accretion disk is present 
during the outbursts.

\end{abstract}


\twocolumn

\section{Introduction}

The pulsar GS 1843-02 was discovered with Ginga in 1988 (\cite{Makino88})
during a galactic plane scan conducted as part of a search for transient
pulsars (\cite{Koyama90b}). These observations found 6 new transient X-ray
sources near galactic longitude $l = 30^\circ$. Their locations and inferred
distance of 10 kpc (based on hydrogen column density estimates)  placed them
within the ``5 kpc arm'',  an inner spiral arm of the galaxy identified with
radio, infrared and  CO molecular line observations (\cite{Hayakawa77}).  GS
1843-02 was detected with a flux of 8 mCrab. Periodic pulses with frequency of
10.5432$\pm$0.0002 mHz were detected, with a pulse profile having a deep narrow
notch seen in all energy bands.  A cross scan located the source to a 50' x 6'
error box (90\% confidence). The SAS-3 source 2S 1845-024 was within the
original Ginga error box (\cite{Makino88}), but is outside of the refined Ginga
error box (\cite{Koyama90a}).

Koyama et al. (1990b\markcite{Koyama90b}) concluded the six new sources they
had discovered in the 5 kpc arm were all transient accreting 
pulsars in Be-star binary systems. Accreting pulsars in binary systems with 
Be (or Oe) stars form the
largest class of known accreting pulsars.  With the exception of X-Perseus,
all of these systems are transients. Be stars are main-sequence stars of
spectral type B that show  Balmer emission lines (see \cite{Slettebak88} for a
review). This line emission  and a strong infrared excess is associated with
the presence of a cool circumstellar envelope.  There is strong observational
evidence that this envelope has the form of a disk, extending along the Be star
equatorial plane  (e.g. \cite{Quirrenbach97}). In Be/neutron star binary systems the
transient  outbursts are thought to be fueled by accretion from this
circumstellar  envelope, frequently occuring near periastron passage. A review
of Be/neutron star binary systems is given by 
Apparao (1994\markcite{Apparao94}).

Zhang et al. (1996) analysed data from the Compton Gamma Ray Observatory (CGRO)
spacecraft's Burst and Transient Source Experiment (BATSE) 
for transient hard X-ray
emission for the region of the sky near $ l = 30^\circ, b=0^\circ$. Using Earth
occultation analysis techniques they identified six
hard X-ray outbursts  between April 1991 and February 1995 from a source,
designated GRO J1849-03, which was localized to a 30' x 60' error box (90\%)
that contained both 2S 1845-024 and portions of the GS 1843-02 error box. The
outbursts were regularly spaced by 241$\pm$1 days, lasted about 13 days, and
reached a peak flux of 75 mCrab (20-100 keV). The outburst behavior suggested a
Be/neutron star binary, with a natural candidate being GS 1843-02. Pulsations
from this source during the outbursts were searched for in the BATSE data, 
but not detected at that time. Observations with
the BeppoSAX wide field camera (\cite{Soffitta98}) during a predicted outburst
of GRO J1849-03 in September of 1996 identified this source with 2S 1845-024. 
Zhang et al. and Soffita et al. suggested the identification the  source with
the transient pulsar GS 1843-02. Pulsations were, however, not detectable in
the BeppoSAX data due to limited sensitivity, leaving this 
conjecture unconfirmed. 

Here we present CGRO/BATSE observations of outbursts of pulsed flux from  GS
1843-02 which are coincident with the outbursts of  GRO J1849-03 = 2S 1845-024
detected in BATSE Earth occultation measurements.  Observations of pulsations
not detected by Zhang et al. (1996) were made possible by sensitivity 
enhancements resulting from improvements in BATSE pulsar analysis techniques 
(see section 2.1). Pulse timing analyses of these observations are presented
which show that the outbursts occur near the periastron passage of a wide,
highly eccentric orbit about a high mass companion. In addition these
observations show that the pulsar spins up during outbursts. We conclude from
the measured spin-up rates and inferred mass accretion rates that an accretion
disk is present during the outbursts. We suggest that the pulsar's orbit is
inclined from its companion's equator, with an outburst initiated and an
accretion disk formed during each passage of the pulsar through its companion's
circumstellar envelope.

\section{Observations and Data Analysis}

BATSE is an all sky monitor designed to study gamma-ray bursts and transient
source outbursts in the hard X-ray and soft gamma-ray bands (\cite{Fishman89}). 
The observations reported here use data from the eight Large Area Detectors
(LADs), which are 2025 cm$^2$ area by 1.24 cm thick NaI (Tl) scintillators
operated in the 20 keV to 1.8 MeV band. These are located at the eight corners
of the CGRO spacecraft. The LADs are uncollimated, with each detector viewing 
half of the sky. The large fields-of-view result in high background rates in
the detectors, which are generally not easily modeled or subtracted. The flux
of a discrete source can be sampled twice per spacecraft orbit by fitting the
steps in count rates that occur when rises or sets over the Earth's horizon
(Earth occultation analysis, see Harmon et al. 1992). For pulsars with periods
shorter than $\sim 500$ s, the variable portion of a pulsar's pulse profile can
be detected by using Fourier analysis, epoch-folding, or by fitting count rates
over short intervals, using a model that includes a description of the pulse
profile, and a polynomial which accounts for the combined background and the
average source flux (pulsar analysis, see Bildsten et al. 1997). 

Our pulse timing analyses use the  DISCLA channel 1
data, which consists of discriminator rates for the  20-50 keV band,
continuously read out from all eight detectors with a  resolution of 1.024 s.
Results from Earth occultation analyses  use the 16 energy channel CONT data,
which is continuously read out from all eight detectors with a resolution of
2.048 s.

\subsection{Blind Frequency Searches}

We have made blind searches for pulsations consistent with the 10.54 mHz pulse
frequency of GS 1843-02 (\cite{Koyama90a}) in the BATSE DISCLA data  from 1991
July 16 to 1997 June 20. Zhang et al. (1996) searched the same data during the
six outbursts of GRO J1849-03 they observed, resulting in no clear  detection
of pulses from GS 1843-02. Through a combination of longer integration times,
optimized combination of rates from different detectors,  and improved handling
of systematic effects due to interfering sources,  our current searches are
approximately 3 times more sensitive then those made  by Zhang et al. (1996).
With these improvements, and the extended search  interval, we clearly detect
10 outbursts from GS 1843-02, at times consistent with the  outburst ephemeris
of GRO J1849-03.  
 
The data were divided into four day intervals, with each interval searched for
pulsations in the barycentric frequency range of 10.47-10.61 mHz. The increase
of the search interval from the one day used by Zhang et al. (1996) to four
days provides a factor of two improvement of sensitivity for a constant
frequency source. No additional increases in the search interval duration 
were made because of the increased danger that frequency changes due to
intrinsic spin-up (or spin-down) and orbital motion would result in a loss of
coherence (and hence sensitivity) in the search (\cite{Chakrabarty97}).

For each four day interval DISCLA channel 1 data was selected for which the
source was visible, the high voltage was on, the spacecraft was outside of the
South Atlantic Anomaly, and no electron precipitation events or other anomalies
had been flagged by the BATSE mission operations personnel. The
rates were combined over detectors using weights optimized for a source with an
exponential spectrum $dN/dE = (A/E)\exp(-E/kT)$ with temperature $kT = 25$ keV
(see Bildsten et al. (1997\markcite{Bildsten97}) appendix A). This weighting
provides an improvement of sensitivity over the cosine of aspect angle
weighting used in Zhang et al. (1996). These combined rates were grouped into
segments of $\approx$300 s duration. This segment duration was chosen because
it contains several pulse periods, and yet is short enough that a quadratic
model generally provides a good fit to the detector background. Segment
boundaries were chosen to avoid inclusion of occultation steps from the bright 
sources Cyg X-1, GRO J1744-28, and GRO J1655-40.

A pulse profile estimate was obtained from each segment by
fitting the combined rates with a model composed of a sixth order Fourier 
expansion in pulse phase (representing the pulse
profile) combined with a quadratic in time (representing the background plus
mean source rate). This resulted in an estimated pulse profile (represented by 6
Fourier coefficients) for each data segment. The phase model used in this
initial fitting had a constant barycentric frequency of $\nu_0$ = 10.54 mHz.
Six harmonics were used so that the narrow notch seen in the Ginga pulse
profiles could be reasonably represented. 

The set of (typically several hundred) 
segment profiles for a given four day interval were 
then searched for a pulsed signal using a generalization of the $Z^2_n$ test 
of Buccheri et al.(1983\markcite{Buccheri83}). 
The $Z^2_n$ test assumes Poisson statistics.
However, the noise in our data is often dominated by the aperiodic activity of
Cygnus X-1, which is usually in the field of view. Our test statistic, which we
call the $Y_n$ statistic, accounts for this frequency dependent excess noise. 

We represent the pulse profile in the $m^{th}$ 300s data segment as   
\begin{equation} 
r(t) =  \Re~\sum_{h=1}^n \alpha_{mh} \exp(i2\pi h\phi_0(t))  
\end{equation} 
where $\Re$ denotes the real part,  $n = 6$ harmonics, $\alpha_{mh}$ is the
estimated complex Fourier coefficient for interval $m$ and harmonic $h$, and 
the initial phase model is $\phi_0(t) = (t-\tau)\nu_0$,  with $t$ the
barycenter corrected observation time and the epoch  $\tau$ = JD245000.5.

We fit the segment profiles in a four day interval to a mean profile which is
represented as
\begin{equation} 
    \bar r(\phi) =  \Re~\sum_{h=1}^n \mu_{h} \exp(i2\pi h \phi)  
\end{equation}
where $\mu_h$ are the estimated Fourier coefficients of the mean pulse profile. 
The pulse phase within segment $m$ is represented as
\begin{equation}
\phi(t) = \phi_0(t)+\Delta \phi_m
\end{equation}
where
\begin{equation}
\Delta\phi_m = ({\bar t}_m - \tau)\Delta \nu~~,
\label{phase_offset}
\end{equation}
with $\bar t_m$ the midpoint
barycentric time of segment $m$, and $\Delta \nu$ a frequency correction
estimated for the four day interval. Here we assume the phase change
within a 300 s segment caused by $\Delta\nu$ is a small fraction of a cycle.

The $\chi^2$ of the fit has the form 
\begin{equation} 
\chi^2 = \sum_{h=1}^n J_h(\mu_h)
\end{equation}
with
\begin{equation}
J_h(\mu_h) = \sum_{m=1}^M
{{|\alpha_{mh}-\mu_h \exp(i2\pi h \Delta\phi_m) |^2} 
 \over {\sigma^2_{mh}}} 
\label{fit_eq}
\end{equation}
where $M$ is the number of 300s segments in the four day interval and
$\sigma_{mh}$ is the Poisson error of either the real or imaginary part 
of $\alpha_{mh}$.
 
The $Z^2_n$ statistic may be expressed as 
\begin{equation} 
Z^2_n ~\equiv~ \sum_{h=1}^n {{|\mu^{min}_h|^2} \over {\sigma^2_{\mu_h}}} 
~=~\sum_{h=1}^n \left[J_h(0)- J_h^{min}\right] 
\end{equation}
were the $J_h^{min}$ is the minimum of $J_h$ which occurs at the mean Fourier
coefficient $\mu_h^{min}$, and $\sigma_{\mu_h}$ is the formal (i.e.
Poisson statistical) error of 
$\mu_h^{min}$.   To account for the non-Poisson errors, 
we multiply $\sigma^2_{\mu_h}$ by the reduced $\chi^2$ 
factor $J_h^{min}/(2M-2)$. This results in
the following test statistic: 
\begin{equation} 
Y_n = \sum_{h=1}^n
\left[J_h(0)-J_h^{min}\right]{{2M-2} \over {J_h^{min}}} 
\end{equation}

\begin{figure}[t!]
\psfig{file=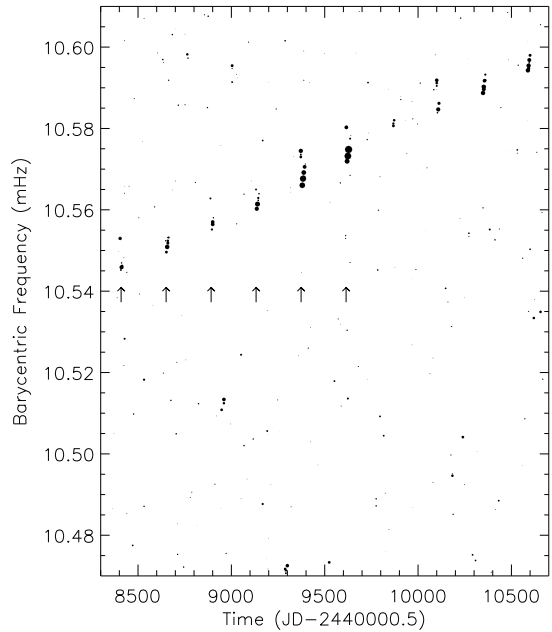,width=3.1in}
\caption{Blind search results for pulsations from GS
1843-02. For each four day search interval the plot shows peaks in the $Y_6$
statistic with value above 35. The symbol area is proportional to $Y_6-35$. The
arrows indicate the peak times of the outburst of GRO J1849-03 reported by
Zhang et al. (1996). For reference JD 2448500.5 = 1991 September 1.0.
\label{raw_freq}}
\end{figure}   

For each four-day search interval the $Y_6$ statistic was calculated for a grid
of frequency offsets. The results are summarized in Fig.  \ref{raw_freq}.
This shows for each search interval the peaks in $Y_6$ with value above 35,
with the symbol size in proportion to $Y_6-35$. Monte-Carlo calculations show
that with no signal present we should expect $\approx$75 peaks on the plot with
$Y_6 > 35$, and a 50\% chance of one peak with $Y_6 > 45$. We actually find
55 with $Y_6 >45$ in Fig. \ref{raw_freq}, with a maximum value 
of $Y_6 = 140$. 

A series of ten outbursts is clearly evident, as is a long-term spin-up trend
in the pulse frequency of  $\approx 2.7\times10^{-13}~{\rm Hz~s}^{-1}$.    The
arrows give the peaks of the outbursts of GRO J1849-03 detected by Zhang et al.
(1996\markcite{Zhang96}) using BATSE occultation techniques. These are
consistent with the pulsed outbursts we have detected. The regular outbursting
is typical of many of the Be star/pulsar systems observed by BATSE
(\cite{Bildsten97}).

\begin{figure}[t!]
\psfig{file=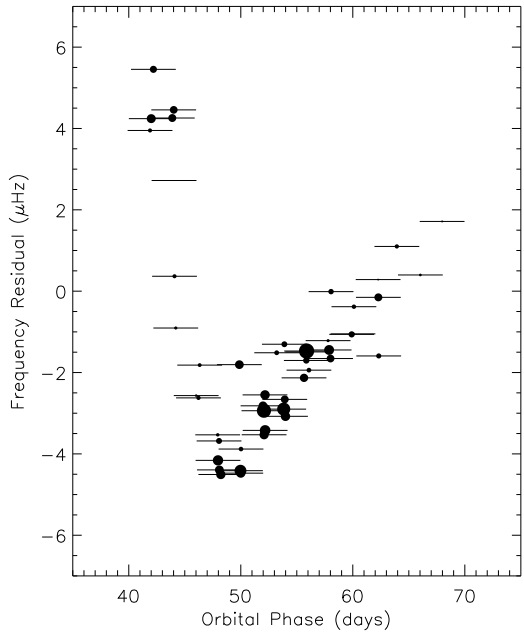,width=3.1in}
\caption{Frequency changes during the outbursts. 
Pulse frequencies were
determined by searches over a narrow frequency band chosen to include the
outbursts seen in Fig. 1. The detected frequencies minus a linear trend
(see text) are plotted against the observation time folded at a 242 day
period, beginning with MJD 48362, which provides an approximate orbital phase.
A highly eccentric binary orbit combined with accretion
induced spin-up caused the observed signature. The periastron epochs given
in Table 1 are at phase 44 days in this plot.
\label{folded_freq}}
\end{figure}

Fig. \ref{folded_freq} examines how the observed pulse frequency changes
during the  outbursts.  For each four day interval we narrowed our frequency
search range to a 14 $\mu$Hz band centered on the linear frequency model
$\nu_{\rm model}(t) = \nu_0+(t-\tau){\dot \nu}$, with  $\nu_0~=~$10.585 mHz,
$\tau~=~{\rm JD}~2450000.5$ and  $\dot \nu ~=~2.72\times10^{-13}{\rm Hz}~{\rm
s}^{-1}$. The frequencies of maximum $Y_6$ are plotted  with this linear model
subtracted if $Y_6 > 35$, with symbol areas proportional to $Y_6-35$. The
abscissa is the observation times folded with a 242 day period, with epoch MJD
48362 (the first day of useful BATSE data). A 242 day period was used rather
than the $241\pm 1$ day orbital period estimate of Zhang et al. (1996) because
it resulted in less scatter. The horizontal bars give the four day width of the
search intervals.  The rapid frequency decrease at the beginning of the
outburst is undoubtedly due to the Doppler signature of an eccentric binary
orbit.  The slower increase in the observed frequency that follows could be due
to both orbital doppler shifts and intrinsic accretion-induced  spin-up of the
pulsar.

\subsection{Pulse Frequency Modeling}

To obtain a preliminary estimate of the system parameters and any intrinsic
spin-up occurring during the outbursts, the frequencies in Fig.
\ref{folded_freq} were fit with a model which included an independent spin
frequency and spin frequency rate for each outburst, and a full set of binary
orbital parameters. A new search of the four day intervals was then conducted,
and new frequency estimates obtained.
In this search the segment phase offsets in equation \ref{phase_offset}
were replaced with 
\begin{equation}
\Delta\phi_m ~=~ \phi_{\rm model}(\bar t_m)+\Delta\nu (\bar t_m-\tau)
-\phi_0(\bar t_m)
\end{equation}
where $\phi_{\rm model}(t)$ encorporates the spin-frequencies, frequency rates, 
and orbital parameters determined by the fit to the initial frequency
estimates.
This iterative step was needed to improve
the sensitivity in intervals where rapid frequency changes were inferred, 
such as
near the beginning of the outbursts.  The search was restricted to frequency
offsets $\Delta \nu$ less than $5.8~\mu$Hz from the new 
folding model. The model parameters were
then re-estimated using the barycentric frequencies determined for detections
with $Y_6 > 35$.

\begin{deluxetable}{rll}
\tablenum{1}
\tablewidth{0pt}
\tablecaption{Orbital Parameter Estimates\label{orbit_tab}}
\tablecolumns{3}
\tablehead{\colhead{} & \colhead{Frequency Fit} & \colhead{Phase Fit}}
\startdata
Period                 &242.18$\pm$0.02 days    & 242.180$\pm$0.012 days      \nl
Periastron epoch       &JD 2449617.3$\pm$0.2    & JD 2449616.98$\pm$0.18      \nl
Semi-major axis        &725$\pm$40 lt-s         & 689$\pm$38 lt-s             \nl
Eccentricity           &0.87$\pm$0.01           & 0.8792$\pm$0.0054           \nl
Argument of periastron &$264^\circ \pm 7^\circ$ & $252.2^\circ \pm 9.4^\circ$ \nl
$\chi^2_\nu$           &926/27                  & 55.1/52                     \nl
\enddata
\end{deluxetable}

\begin{figure}[b!]
\psfig{file=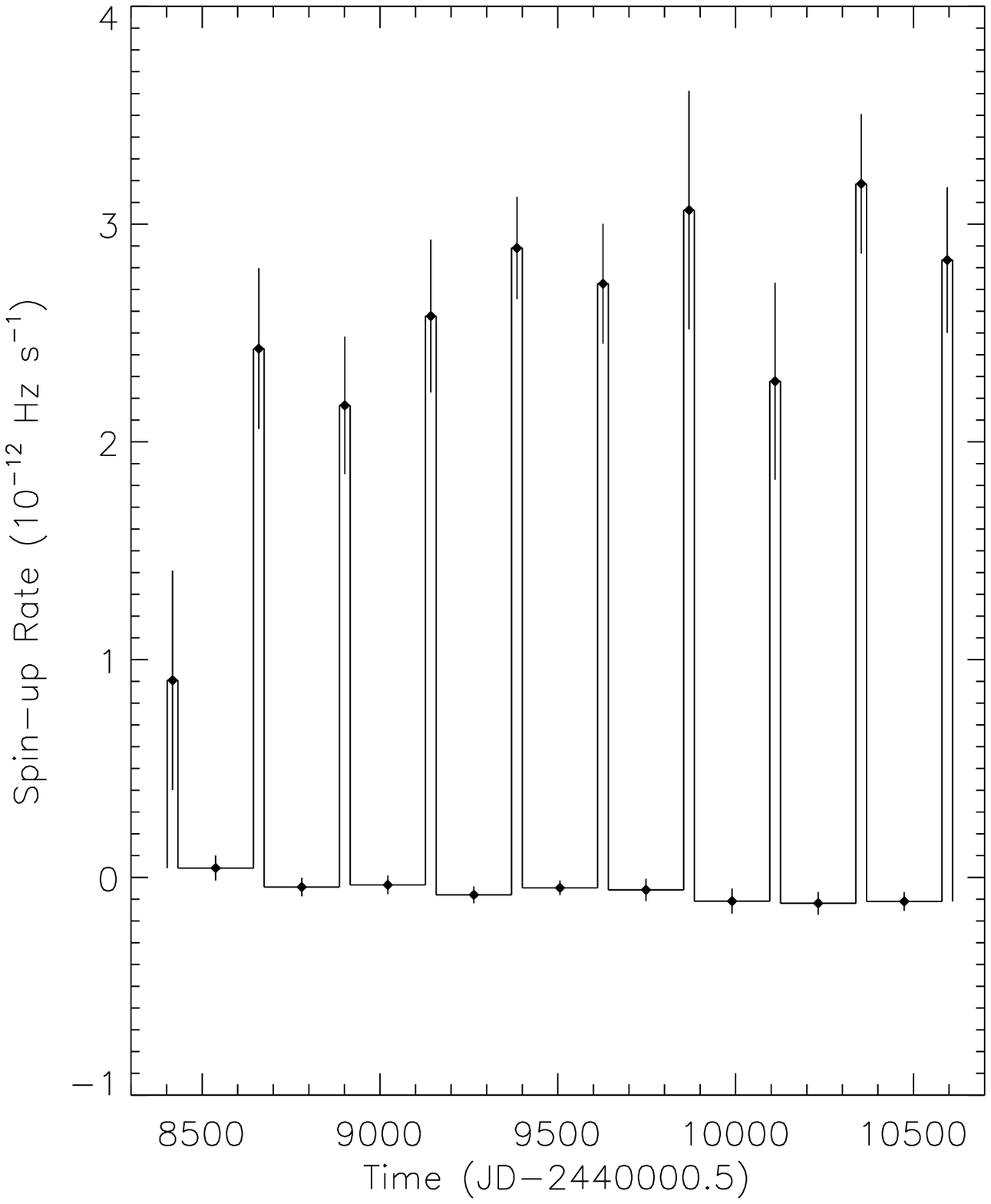,width=3.1in}
\caption{The intrinsic spin-up rates estimated by fitting the pulse
frequencies. The model assumes the spin-up rate is constant for 30 day
intervals containing each outburst. The rate between outbursts is calculated
from the
model frequency difference between the edges of the adjacent outburst
intervals. \label{freq_fdot}}
\end{figure}

The first column of Table \ref{orbit_tab} shows the estimated binary orbital 
parameters, while
the spin-up rates during the outbursts, and those inferred between outbursts
(using a 30 day outburst width) are given in Fig. \ref{freq_fdot}. The highly
eccentric orbit has a mass function of $7\pm 1 M_\odot$, suggesting a $9
M_\odot$ companion or larger. The outbursts occur close in phase to the
estimated periastron passage. The fit suggests that the long-term spin-up
trend is the result of accretion torques that occur during the outbursts.

The quality of this fit is poor, with $\chi^2_\nu = 34.3$. The fit residuals
for most of the outbursts show significant structure, indicating the simple
constant frequency rate model used is inadequate. The errors given in Table
\ref{orbit_tab} have been increased by the factor  $\sqrt{\chi^2_\nu}$ to
account for this poor fit, however they are undoubtedly still underestimated.
Adequate modeling will require parameters describing the time history of the
torque during the outburst. These parameters will be strongly coupled to the
orbital elements due to the limited orbital phase coverage. Such modeling will
be studied in the pulse timing analysis that follows. 

While statistically a poor fit, this model is sufficiently accurate for 
coherent epoch folding over intervals of a few days. 
The R.M.S. deviation of the measured frequencies from the model is 
0.21 $\mu$Hz. 

\subsection{Pulse Profile Measurements}

\begin{figure}[b!]
\psfig{file=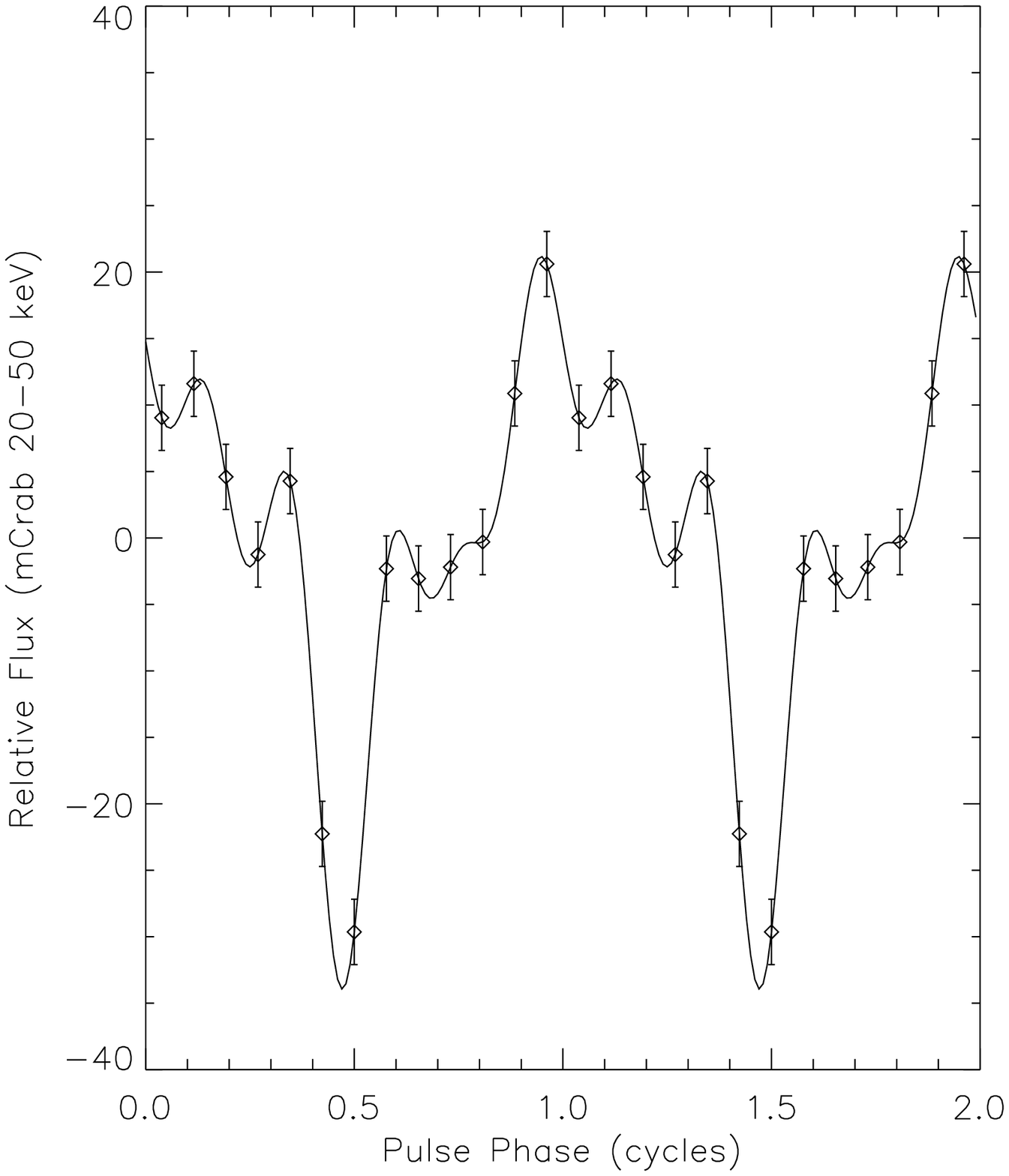,width=3.1in}
\caption{The pulse profile of GS 1843-02 obtained from the BATSE
20-50 keV rates. The profile is limited to six harmonics. Error bars are give
at approximately statistically independent points.
\label{profile}}
\end{figure}

Fig. \ref{profile} shows the mean pulse profile from the most significant
four day interval (MJD 49624.0-49628.0).  The profile is derived from estimates
of its first 6 Fourier amplitudes, using the frequency model discussed in
section 2.2, and is therefore a smooth curve. Errors are shown at phases which
are approximately statistically independent.  The pulse amplitude is
$\approx$20 mCrab R.M.S. (20-50 keV). This profile shows a narrow notch, which
is similar to the GS 1843-02  profile observed by Ginga in the 9-38 keV range
(\cite{Koyama90a}).  

In preparation for a pulse timing analysis, one mean pulse profile was
estimated
every two days during the time range of each outburst for which pulsations 
were detected. Two profiles were made for each four day frequency search
interval to provide higher resolution during the periastron passage. 
We found that profiles from one day integrations were often not of sufficient
significance to reliably estimate pulse phases.
The first 6 Fourier amplitudes
of each mean pulse profile were estimated from the 300s profiles using the
phase model obtained from the fit to pulse frequencies discussed above. The
Poisson statistical errors of each Fourier amplitude $\mu_k$ was modified by
the factor $(J_h(\mu_h)/[2M-2])^{\onehalf}$, were M is the number of 300s
profiles, to account for non-Poisson noise.  Several profiles from the edges of
outbursts were dropped from further analysis because of low significance. 

\subsection{Pulse Phase Measurements}

A phase offset $\Delta \phi$ from the phase model was estimated for 
each two-day mean pulse profile by 
fitting the Fourier amplitudes $\mu_h$ of the
profile to the Fourier amplitudes of a scaled and shifted template profile: 
\begin{equation}
\chi^2 = \sum_{h=1}^6 {{| \mu_h - A\exp(i2\pi\Delta\phi)T_h|^2} 
       \over {\sigma^2_{\mu_h}}}~.
\label{template_fit}
\end{equation}
Here $A$ is the estimated pulse amplitude, $T_h$ is a Fourier amplitude of
the template profile, and $\sigma_{\mu_h}$ is the error on the real or imaginary
component of $\mu_h$.
 
The template profile was obtained in three steps. 
Initially a cosine template was
used to obtain phase offsets in the bright interval MJD 49620.0-49630.0. The
profiles in this interval were then aligned using these offsets and a new
template constructed from the weighted average of the aligned profiles. Using
this template, phase offsets and amplitudes were than estimated for all the
profiles. The 46 profiles with pulse amplitude above 5$\sigma$ significance
where than aligned using these estimated offsets, and the final template
constructed from the weighted average of these aligned profiles.

For each two-day interval $j$ the mean observation time, reduced to the solar 
system barycenter, $t_j$, was computed. The total pulse phase at that
time was then computed as $\phi_j = \phi_{\rm model}(t_j)+\Delta \phi_j$,
where $\Delta \phi_j$ is the estimated phase offset in interval $j$.

\subsection{Pulse Phase Modeling}

The pulse phases were fitted with a model that assumed a Gaussian profile for
the spin-up rate during each outbursts. In outburst $k$ the spin-up rate
is described by three adjustable parameters; the peak spin-up rate $\dot\nu_k$, 
time of peak $t^{\rm peak}_k$, and outburst width $W_k$.
During outburst $k$ the fitted phase model had the form
\begin{eqnarray}
&\phi_{\rm model}(t) ~=~ \psi_k+\nu_k(t^{\rm em}-\tau_k) &\\
    &+\dot \nu^{\rm peak}_k
     \int_{t^{\rm peak}_k}^{t^{\rm em}}\int_{t^{\rm peak}_k}^{\hat{t}} 
     \exp\left(-{\onehalf}{{(\tilde{t}-t^{\rm peak}_k)^2} \over {W^2_k}}\right)
      d\tilde{t}d\hat{t}&.\nonumber
\end{eqnarray}
Here the phase constant $\psi_k$ and the
frequency constant $\nu_k$ 
are adjustable parameters, while the epoch $\tau_k$ is fixed near the 
center of the outburst. The emission time $t^{\rm em}$ is related to the
barycenter corrected observation time $t$ by
\begin{eqnarray}
t & = &  t^{\rm em}+a_x\sin i(\sin\omega [cos E-e] \\
  && +\sqrt{1-e^2}\cos\omega\sin E)/c \nonumber
\end{eqnarray}
where
\begin{equation}
E - e\sin E = {{2\pi} \over {P_{\rm orbit}}} (t^{\rm em}-\tau_{\rm periastron})
\end{equation}
with the orbit's projected semi-major axis $a_x\sin i$, argument of
periastron $\omega$, eccentricity $e$, orbital period $P_{\rm orbit}$, 
and periastron epoch $\tau_{\rm periastron}$, all being adjusted in the fit.
The fitting was performed by minimizing
\begin{equation}
\chi^2 = \sum_{j=1}^N {{[\phi_k-\phi_{\rm model}(t_k)]^2} \over 
{\sigma_{\phi_k}^2}}
\end{equation} 
using the Levenberg-Marquardt method (\cite{NumRecipes}). 

\begin{figure}[t!]
\psfig{file=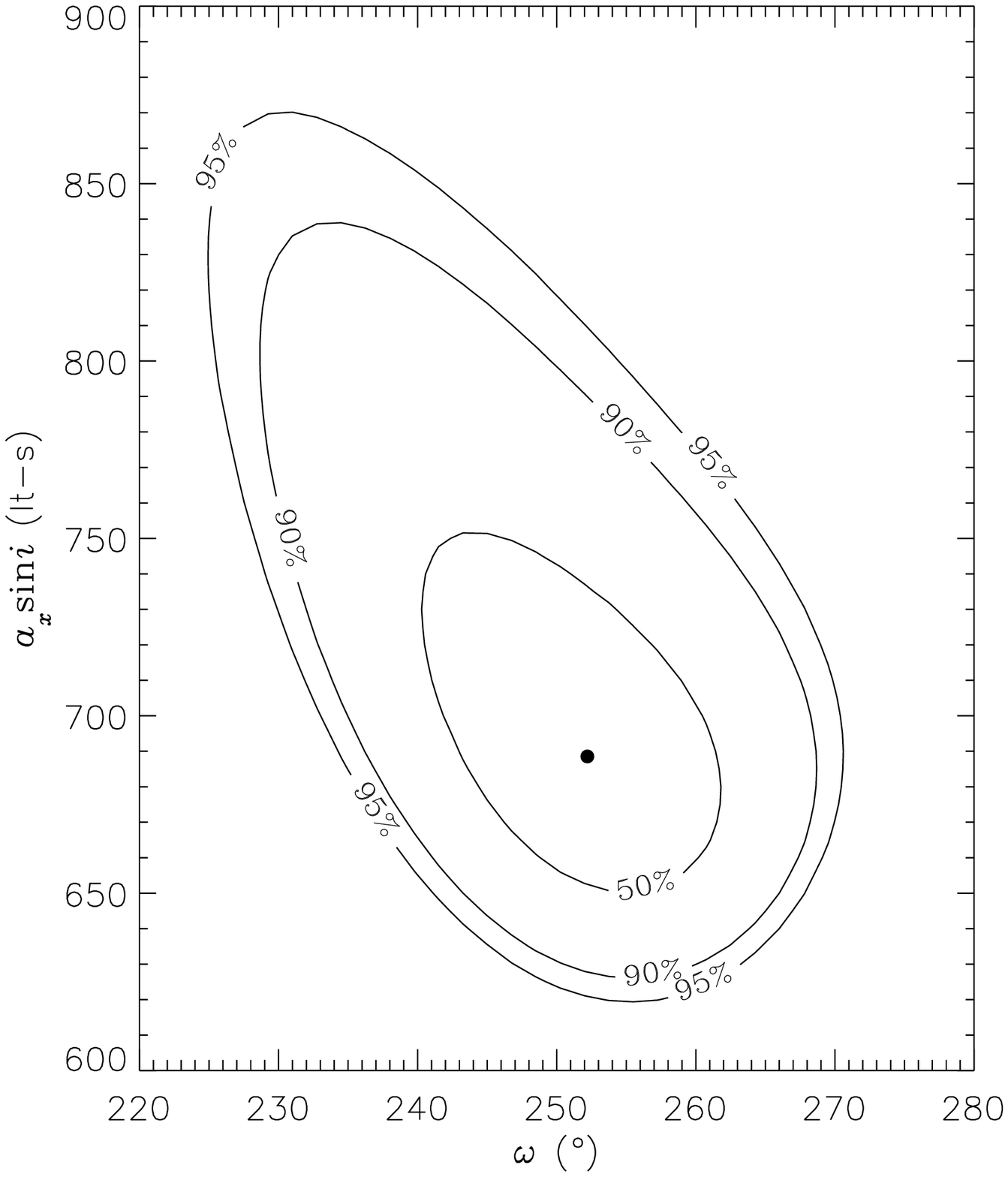,width=3.1in}
\caption{Joint confidence regions for the argument of periastron
$\omega$ and the projected semi-major axis $a_x\sin i$ as estimated by fitting
pulse phases.
\label{omega_ax}}
\end{figure}

Initially only a single common value was estimated for the widths $W_k$, and
the times of peak $\tau_k$ were constrained to be separated by the orbital
period. This fit had a $\chi^2$ of 135.4 with 70 degrees of freedom. Allowing
the $\tau_k$ to be independently adjusted reduced the  $\chi^2$ to 74.7 with 61
degrees of freedom. With the widths $W_k$ also independently adjusted, $\chi^2$
dropped to 55.1 with 52 degrees of freedom. 

The orbital parameters obtained from this fit are shown in the second column
of Table \ref{orbit_tab}. These are consistent with those 
obtained by fitting observed frequencies.  The period,
periastron epoch, and eccentricity are all well constrained by the pulse phase
measurements. The argument of periastron $\omega$ and the semi-major 
axis $a_x \sin i$ are not as well constrained.  
Fig. \ref{omega_ax} shows their joint
parameter confidence region. The broad minimum
of $\chi^2$ for these two parameters is due to the difficulty in separating
orbital and intrinsic torque signatures with data restricted to a small
interval of orbital phase.

\begin{figure}[t]
\psfig{file=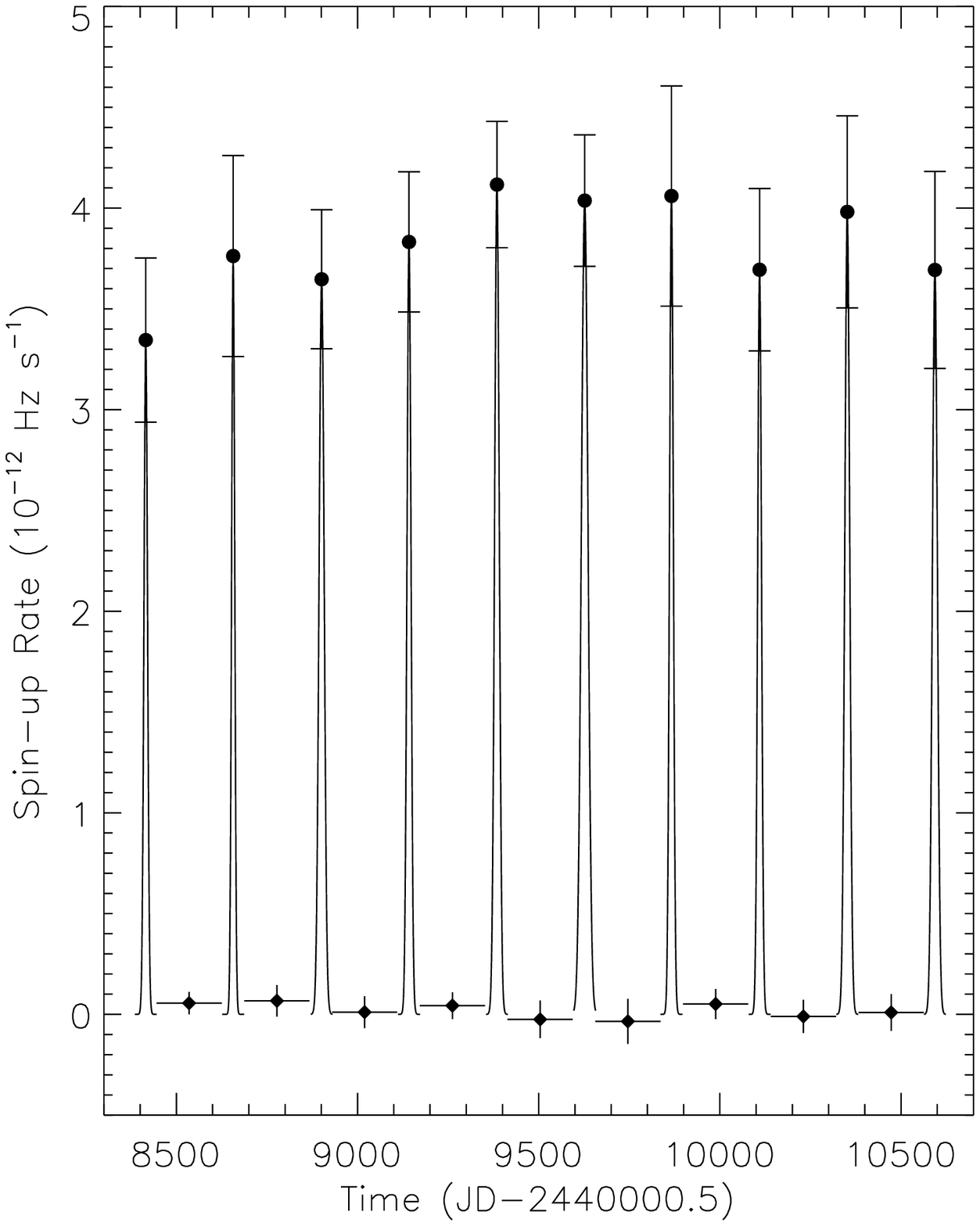,width=3.1in}
\caption{The intrinsic 
spin-up rates estimated by fitting the pulse
phase. The model profile is Gaussian with adjustable width and peak time for
each outburst. The solid circles give the peak spin-up rates. The average
spin-up rates between outbursts are given by the solid diamonds. These are
calculated from the estimated model for 
182 day intervals centered between outbursts.
\label{nudot2}}
\end{figure}

The estimated spin-up rate model is shown in Fig. \ref{nudot2} which 
also shows the mean spin-up rate inferred between outbursts. Torques between
outbursts contributing an estimated $5\pm23$\% of the long term frequency
change. The error on this estimate is coupled to those of $\omega$ and
$a_x \sin i$, with the estimated fraction varying from -0.33 to 0.26 for
fits with fixed values of $\omega$ and $a_x \sin i$ within the 50\% confidence
contour shown in Fig. \ref{omega_ax}.

\begin{figure}[t!]
\psfig{file=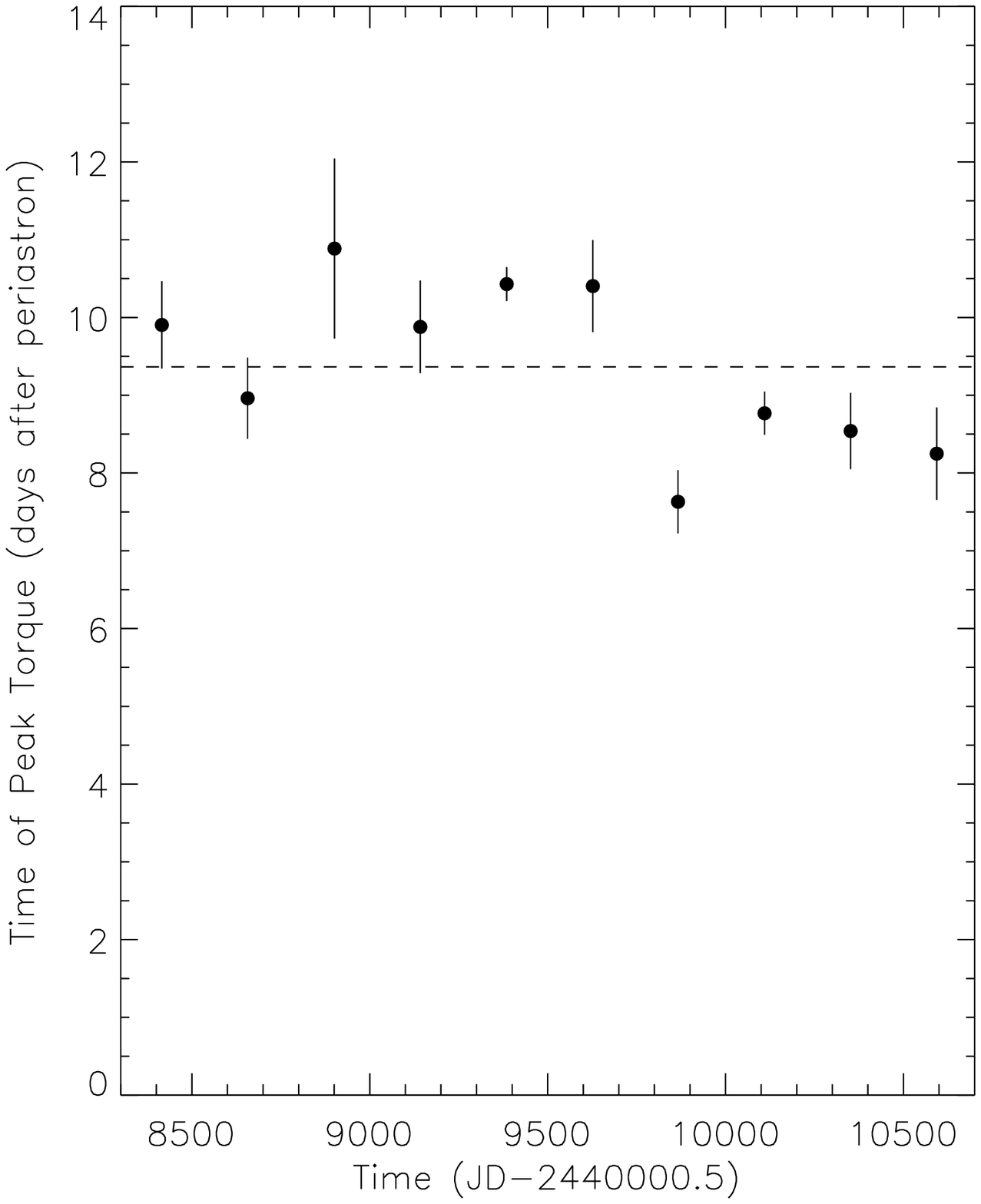,width=3.1in}
\caption{The times of peak spin-up rate $t^{\rm peak}_k$ 
relative to periastron passage estimated from the pulse phase fit.
The errors shown are for the difference from the mean delay of 9.4$\pm$0.7 days,
which is shown by the dashed line.
\label{tpeak}}
\end{figure}

Fig. \ref{tpeak} shows the estimated delay of the spin-up rate peak from
periastron passage. The mean delay is 9.4$\pm$0.7 days, which is shown in the
figure. The error on the mean delay is coupled to those of $\omega$ and $a_x
\sin i$, with the estimated mean delay varying from 8.1 to 10.0 days for fits
with fixed values of $\omega$ and $a_x \sin i$ within the 50\% confidence
contour shown in Fig. \ref{omega_ax}. The errors on the individual delays are
correlated due to this coupling between orbital and torque parameter errors. In
order to show the significant differences between delays, the error shown for
each delay is that of the difference of the delay from the mean delay.

\begin{figure}[t!]
\psfig{file=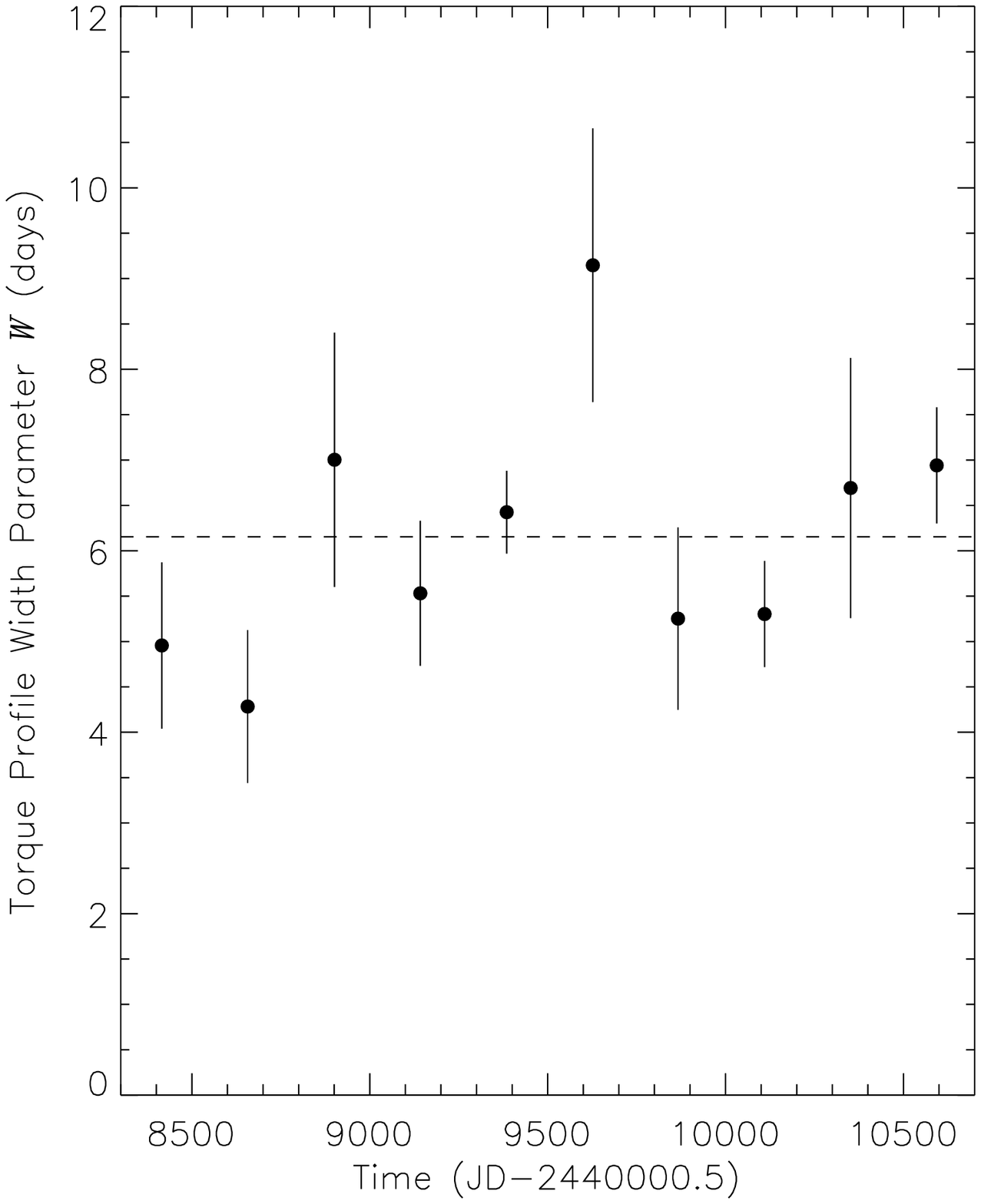,width=3.1in}
\caption{The estimated widths $W_k$ of the spin-up rate
profiles. The errors shown are for the difference from the mean of 6.2$\pm$1.0
days, which is shown by the dashed line.
\label{width}}
\end{figure}

Fig. \ref{width} shows the estimated width parameters $W_k$  of the spin-up
rate profiles. This has a mean value $6.2\pm1.0$ days, which is shown in the
figure. The error on the mean width is coupled to those of $\omega$
and $a_x \sin i$, with the estimated mean varying from 5.2 to 7.6 days for
fits with fixed values of $\omega$ and $a_x \sin i$ within the 50\% confidence
contour shown in Fig. \ref{omega_ax}. The errors on the individual widths are
correlated due to this coupling between orbital and torque parameter errors. In
order to show the significant differences between widths, the error shown for
each width is that of the difference of the width from the mean width.

\subsection{Outburst Profile}

\begin{figure}[t!]
\psfig{file=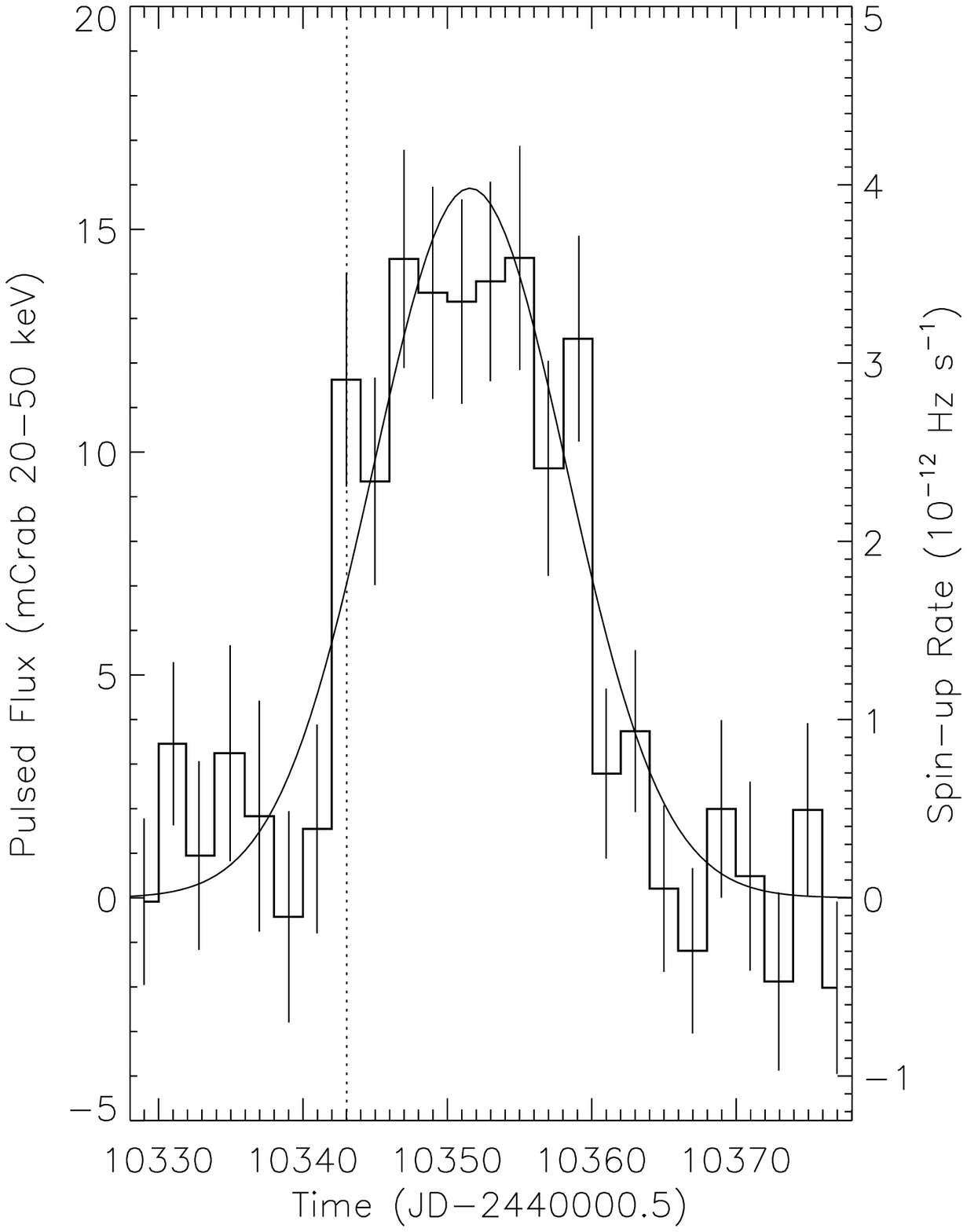,width=3.1in}
\caption{The 20-50 keV pulsed flux (R.M.S. normalized) 
during the 1996 September outburst. 
Also shown is the model torque profile for this outburst (smooth
curve). The dotted line is the time of periastron passage.
\label{outburst8}}
\end{figure}

Fig. \ref{outburst8} shows the lightcurve of the pulsed flux for the 1996
September outburst. The other outbursts are similar. A mean pulse profile was
determined from two day intervals, using the detailed phase model estimated in
the previous section  to align the 300s segment profiles. Non-Poisson noise was
accounted for in the errors as previously described. The pulsed flux for each
interval was calculated from the correlation of the mean pulse profile in that
interval with the pulse template (with no adjustment in phase). The template
was normalized to have unit variance, so that the correlation gives the R.M.S.
pulsed flux if the pulse profile has the same shape as the template. Also shown
in the plot is the model torque profile for the outburst, and the time of
periastron passage. 

\begin{figure}[t!]
\psfig{file=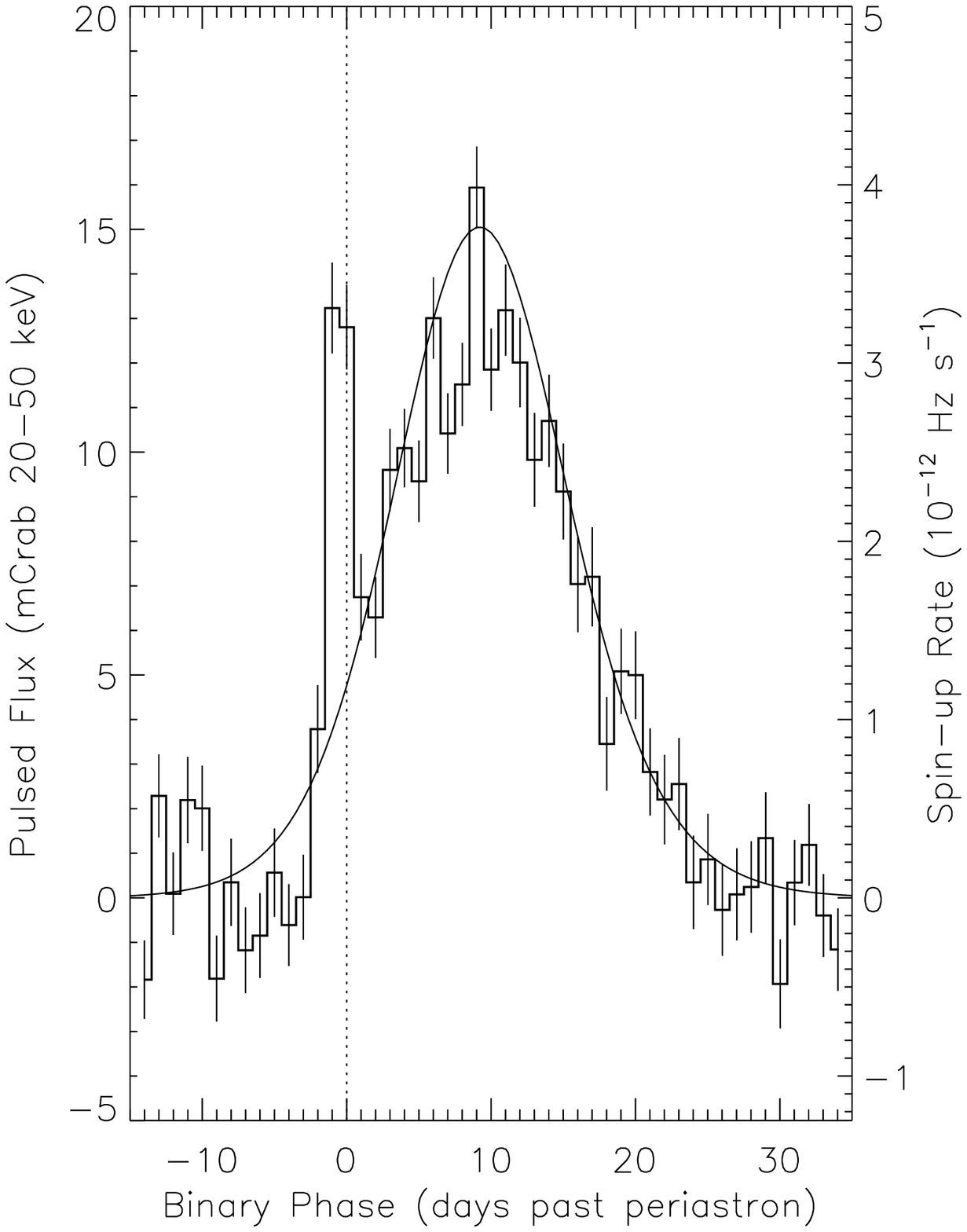,width=3.1in}
\caption{The 20-50 keV pulsed flux (R.M.S. normalized)
epoch folded at the orbital period. Also shown is the average model torque 
profile.
\label{avgoutburst}}
\end{figure}

To examine the outburst lightcurve with higher signifigance an average 
outburst profile was constructed. This is shown in Fig. \ref{avgoutburst}.  
The pulsed flux was determined as above for $P_{\rm orbit}/242 \approx 1$ day
intervals and then epoch folded at the orbital period $P_{\rm orbit}$. Also shown
in the figure is the average model torque profile. Overall the pulsed flux and
the torque profile are well aligned and of approximately the same width. Since
the widths and peak times of the torque profile for each outburst were
estimated in the phase modeling, this demonstrates a correlation between the
spin-up rate and the pulsed flux. 

There is significant narrow feature in the average outburst lightcurve near
periastron passage. This has a width of two days or less, and centered
approximately a half day before periastron. Inspection of the 10 individual
outburst lightcurves at 1 day resolution shows that this feature is present 
in the majority of them, and is not caused by any individual flux enhancement.
 
\begin{figure}[t!]
\psfig{file=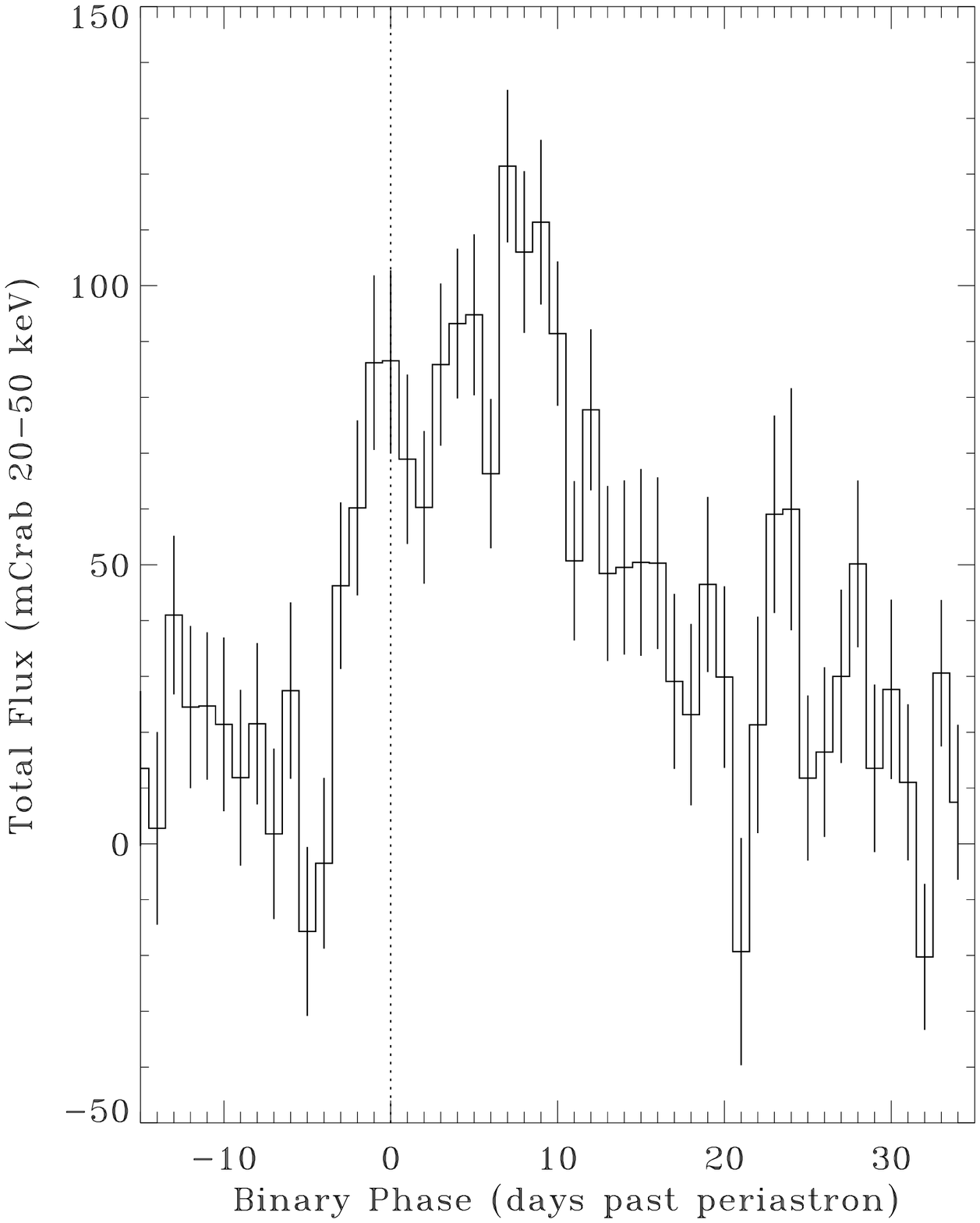,width=3.1in}
\caption{The 20-50 keV total flux (from BATSE 
Earth occultation measurements) epoch folded at the orbital period.
\label{folded_occ}}
\end{figure}

In figure \ref{folded_occ} we show the average total flux profile obtained by
epoch folding, at the orbital period, fluxes obtained by Earth occultation
analysis (\cite{Harmon92}) of the BATSE CONT data. This figure includes only
the outbursts analysed by Zhang et al. (1996) using data from MJD 48363-49741.
The occultation fluxes were obtained from fits that used a power-law spectral
model with a fixed photon number index of -2.5. The onset time and width of the
outburst roughly agrees with that in figure \ref{avgoutburst}, but a detailed
comparison is precluded by the higher relative noise level in the average total
flux profile.

\section{Discussion}

These observations show a series of 10 outbursts of the pulsar GS 1843-02
occurring every 242 days. This outbursting behavior, and the wide,  highly
eccentric orbit, which has a mass function requiring a companion mass
of $M_{opt} > 7 M_\odot$, leave little doubt that the source is a
Be star/pulsar system. 
The consistency of the outburst intervals with those of the
periodic transient GRO J1849-03 (\cite{Zhang96}) clearly shows these to be the
same source. BeppoSAX Wide Field Camera observations of GRO 1849-03 during a
predicted outburst in September 1996 have recently identified this source with 
2S 1845-024 (\cite{Soffitta98}). 2S 1845-024 was observed by SAS 3 in July and
September 1975, and localized to 35" (\cite{Doxsey77}). As shown in Soffitta et
al. (1998) the recent outbursts were also seen in the RXTE ASM lightcurve of 2S
1845-024. 

Other observations of outbursts and additional identifications might possibly
be found by examining sources cataloged in the region of 2S 1845-024 by early 
missions. This includes A 1845-02 and H 1845-024. A 1845-02, which was
discovered by Ariel V in 1974  November (\cite{Villa76}), and was identified
with 2S 1845-024 by Doxsey et al. \markcite{Doxsey77}(1977).  H 1845-024 is
listed in the HEAO 1 catalog (\cite{Wood84})  with the assumed location of 2S
1845-024. In Table \ref{obs_phases} we list observations of 2S 1845-024 and its
possible counterparts along with the calculated orbit phase during the
observations. The detections by Ariel V, SAS 3, Ginga, BeppoSAX, and RXTE  are
all consistent with the range of orbital phases in which  BATSE detects
pulsations, which is from 4 days before to 24 days  after periastron passage.
The HEAO observation however occurs near apastron. This detection is possibly
due to source confusion with another transient. Several other transient 
pulsars that have been found in this region (\cite{Koyama90b}), since the HEAO
1 catalog analysis.

\begin{figure}[t]
\psfig{file=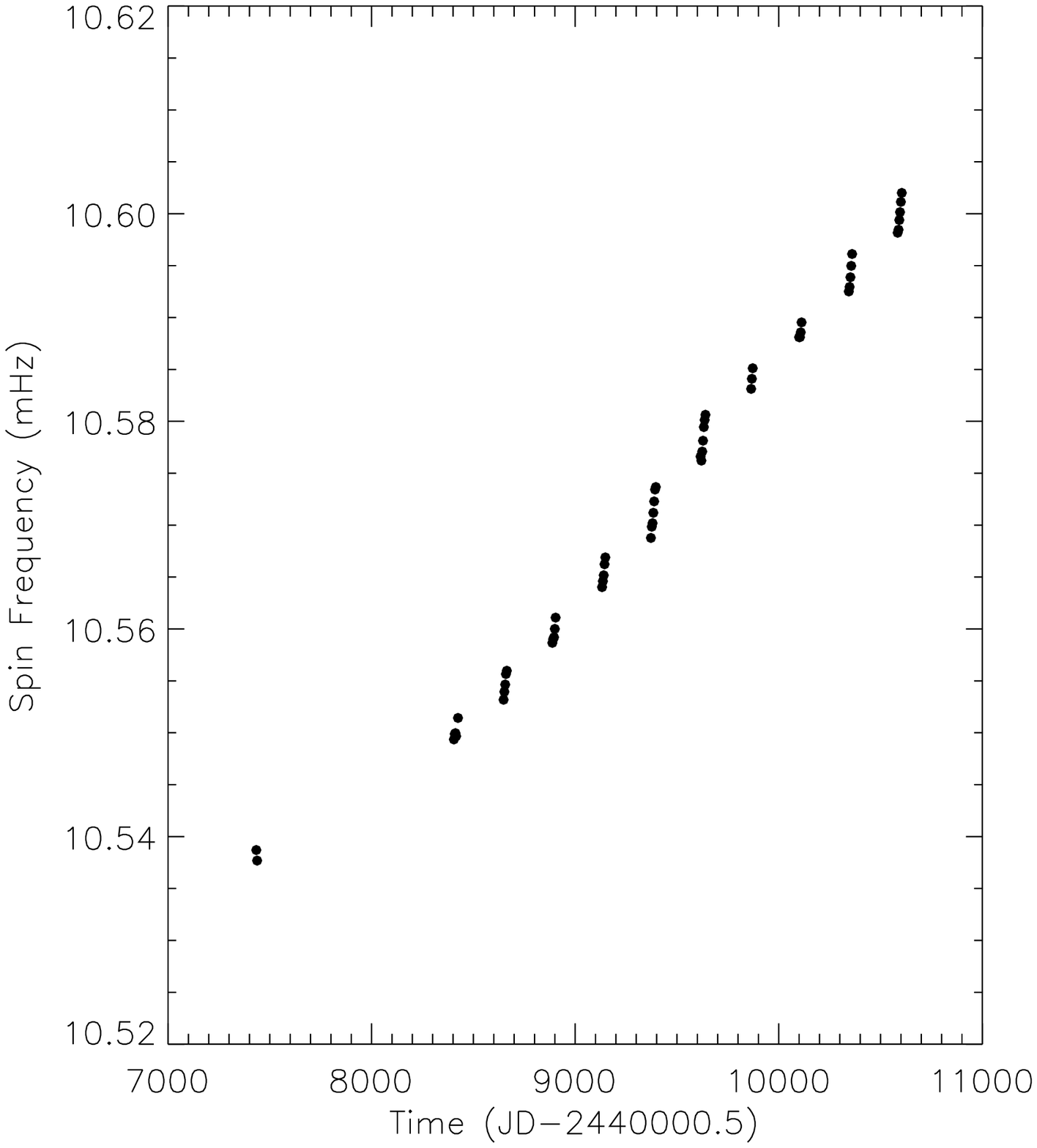,width=3.1in}
\caption{The long-term history of the spin frequency, corrected
using the orbit in Table 2. The initial points are from the Ginga pointed
observations, the remainder from BATSE observations.
\label{lt_freq}}
\end{figure}

BATSE observations of accreting pulsars (\cite{Bildsten97})  have shown that
extended series of low luminosity ($10^{36-37}~{\rm erg}~{\rm s}^{-1}$)
outbursts occurring near periastron are common in the transient 
Be star/pulsar systems. The outbursts of 2S 1845-024 observed by 
BATSE fit into this 
``normal'' outbursting pattern. The peak luminosity determined for the 1996 
September outburst from combined RXTE ASM (2-12 keV) and 
BATSE occultation data (20-100 keV) was $6
\times 10^{36} {\rm erg}~{\rm s}^{-1}$, assuming a distance of 10 kpc
(\cite{Soffitta98}). As shown in Table \ref{obs_phases} the peak fluxes of the
outbursts observed by BATSE are all fairly similar. The outbursts have
proceeded like clockwork since at least the launch of the Compton Observatory.
Fig. \ref{lt_freq} shows the the Ginga and BATSE frequency
measurements, corrected for the orbit in the second column of Table
\ref{orbit_tab}. Significant spin-up occurred between the Ginga and
and the BATSE measurements, but that spin-up was smaller than would be
predicted from an extrapolation of the BATSE measurements.
If, as suggested by our estimates, the long-term spin-up is due largely to 
torques occurring within outbursts, then
unobserved outbursts must have occurred between the Ginga and BATSE
observations. The lower spin-up rate implies that these outbursts were on
average smaller, or that one or more periastron passages occurred without an
outburst.  

\begin{deluxetable}{llcc}
\tablenum{2}
\tablewidth{0pt}
\tablecaption{Orbital Phase of Observations\label{obs_phases}}
\tablecolumns{4}
\tablehead{
\multicolumn{2}{c}{Date} & 
\colhead{Phase} & 
\colhead{Peak Flux} \\ 
\multicolumn{2}{c}{} &
\colhead{(days past periastron)} & 
\colhead{(mCrab)}}
\startdata
\multicolumn{3}{l}{Ariel V\tablenotemark{a}} & (2--6 keV)\nl
&1974 Nov. 12 -- 25    & 11 -- 26 & 12  \nl
&1975 July 21 -- Aug. 16 & 21 -- 48 & 2 \nl
\tablevspace{1pt}
\multicolumn{3}{l}{SAS 3\tablenotemark{b}} &(2--11 keV)\nl
&1975 July 11 -- 16        & 11 -- 17 & 9  \nl
\tablevspace{1pt}
\multicolumn{3}{l}{HEAO 1\tablenotemark{c}} & (2--10 keV)\nl
&1977 Oct. 2 -- 10      & 98 -- 106 & 7 \nl
\tablevspace{1pt}
\multicolumn{3}{l}{Ginga LAC\tablenotemark{d}} & (2--20 keV)\nl
&1987 Oct. 6            & -120      & $<$ 0.25 \nl
&1987 Oct. 10           & -116      & $<$ 0.25 \nl
&1988 Sept. 12         & -20       & $<$ 0.25 \nl
&1988 Sept. 29         & -3        &   8 \nl
&1988 Oct. 3            & 1         &  9 \nl
\tablevspace{1pt}
\multicolumn{3}{l}{BATSE} & (pulsed, 20--50 keV)  \nl
&1991 May 28 -- June 16          &    -2 -- 18  & $18\pm5$  \nl
&1992 Jan. 25 -- Feb. 13  &    -2 -- 18  & $12\pm2$ \nl
&1992 Sept. 21 -- Oct. 12 &    -4 -- 18  & $14\pm2$ \nl
&1993 May 23 -- June 11          &    -2 -- 18  & $15\pm2$ \nl
&1994 Jan. 18 -- Feb. 14  &    -4 -- 24  & $15\pm2$ \nl
&1994 Sept. 21 -- Oct. 12 &    -1 -- 22  & $14\pm2$ \nl
&1995 May 19 -- June 7           &    -3 -- 17  & $20\pm4$ \nl
&1996 Jan. 16 -- Feb. 8   &    -3 -- 21  & $13\pm2$ \nl
&1996 Sept. 16 -- Oct. 5  &    -1 -- 19  & $14\pm3$ \nl
&1997 May 14 -- June 8           &    -3 -- 23  & $15\pm2$ \nl
\tablevspace{1pt}
\multicolumn{3}{l}{BeppoSAX Wide Field Camera\tablenotemark{e}} &(5--21 keV)\nl
&1996 Sept. 17-18            &     0 -- 1   &  9  \nl
\tablevspace{1pt}
\multicolumn{3}{l}
{Rossi X-ray Timing Explorer All Sky Monitor\tablenotemark{f}}&(2--12 keV)\nl
&1996 Sept. 22 -- Oct. 3  & 5 -- 17 & $21\pm 2$ \nl
&1997 May 20 -- May 27  & 3 -- 11 & $22 \pm 2$ \nl
\enddata                                    
\tablenotetext{a}{A 1845-02 was detected in scans folded over the given time
ranges (Villa et al. 1976). Fluxes were calculated from the HE count rates.
Cross scans are mentioned, but accurate dates are not given.} 
\tablenotetext{b}{From Doxsey et al. (1977).}
\tablenotetext{c}{From Wood et al. (1984). The observation interval was
calculated from the ecliptic longitude of 2S 1845-024.}
\tablenotetext{d}{From Koyama et al. (1990a). The fluxes were calculated 
from the 2-20 keV count rates.}
\tablenotetext{e}{From Soffitta et al. (1998).}
\tablenotetext{f}{From 4 day averages using 1 Crab = 75 c\,s$^{-1}$.}
\end{deluxetable}                                  

The BATSE observations of 2S 1845-024 provide the best evidence to date of
spin-up occuring during normal outbursts of a transient pulsar. Spin-up during
the less common ``giant'' outbursts of transient pulsars has been reported many
times. During giant outbursts, which peak near the Eddington luminosity, an
accretion disk is known to be present (\cite{Finger96}). Only recently, however,
has evidence for spin-up during normal outbursts emerged with BATSE
observations of 2S 1417-624 (\cite{Finger96b}),  GS 0834-430
(\cite{WilsonC97}), and EXO 2030+375 (\cite{Stollberg99}), and now the
observations of 2S 1845-024 present here. The most common explanation for
normal outbursting activity in 
transient Be star/pulsar systems is the direct accretion of
the dense  equatorial wind (or circumstellar disk) of the Be star
(\cite{Waters89}) as the neutron star passes through periastron. Yet wind
accretion is thought to be very inefficient at transfering angular momentum
(\cite{Ruffert97}), leading us to expect any spin-up during normal outbursts to
be fairly small. 

The spin-up rate we see during outbursts of 2S 1845-024, and its correlation
with the pulsed flux (Fig. \ref{avgoutburst}) suggests an accretion disk is
present.  A disk will form if the specific angular momentum $l$ 
in a wind exceeds the Keplerian specific angular momentum 
$l_{\rm m} = (GMr_{\rm m})^{\onehalf}$ 
at the magnetospheric radius $r_{\rm m}$. 
After an accretion disk is formed, the specific
angular momentum of accreting material is maintained near $l_{\rm m}$.
For accretion to occur, the magnetospheric radius must be
within the co-rotation radius 
$r_{\rm co} = (GM)^{\onethird}(2\pi \nu)^{-{\twothirds}}$.
Therefore the specific angular momentum needed for disk formation, $l_{\rm m}$,
is bound by  
\begin{equation} 
l_{\rm m} < l_{\rm co} = (GM r_{\rm co})^{\onehalf} =  8
\times 10^{17} {\rm cm}^2 {\rm s}^{-1}~~. \label{disk_formation} 
\end{equation} 
From the luminosity and spin-up rate at the peak of the 1996
September outburst (which is typical) 
we infer a specific angular momentum of the accreting material of 
\begin{equation} 
l = 2 \pi I \dot \nu \dot M^{-1}  = 8 \times
10^{17} d_{10}^{-2} {\rm cm}^2 {\rm s}^{-1}  \label{lmeas} 
\end{equation}
assuming $\dot M = L (GM/R)^{-1}$.  Here $\dot\nu = 4\times 10^{-12}~{\rm
Hz}~{\rm s}^{-1}$ is the spin-up rate,  $\dot M$ is the mass accretion rate, 
$L=6\times 10^{36}~{\rm erg}~{\rm s}^{-1}~d^2_{10}$ is 
the luminosity,  $d_{10}$
is the distance in units of 10 kpc. We have  assumed $I = 10^{45} {\rm
g~cm}^2$ for the neutron star  moment of inertia, $M = 1.4M_\odot$ for the
mass, and $R = 10^6 {\rm cm}$ for the radius.  From comparison of equations
\ref{disk_formation} and \ref{lmeas} we conclude that disk accretion is
occuring during the outbursts of 2S 1845-024.  By moving the source farther
away we could avoid this conclusion. The distance of 10 kpc estimated by Koyama
et al.\markcite{Koyama90a} (1990a) was, however, based on the minimum column
density determined for a set of Ginga spectra. Since part of this column
density may be intrinsic to the source, a closer distance is more likely than a
farther one.   

The progenitor system of a Be/neutron star binary 
is likely a compact binary system of
B stars (\cite{vdHeuvel94}). The more massive star evolves more rapidly and
transfers mass to its companion via Roche-lobe overflow, resulting in a helium
star in a circular orbit about its mass inhanced companion.  The helium star
then undergoes a supernova explosion producing a neutron star in a wide
eccentric orbit, if the system remains bound.  Because the kick velocities
produced by the supernova are thought to be larger than or comparable to typical
orbital velocities of the pre-supernova systems, many such systems are expected
to be disrupted.

\begin{figure}[b!]
\psfig{file=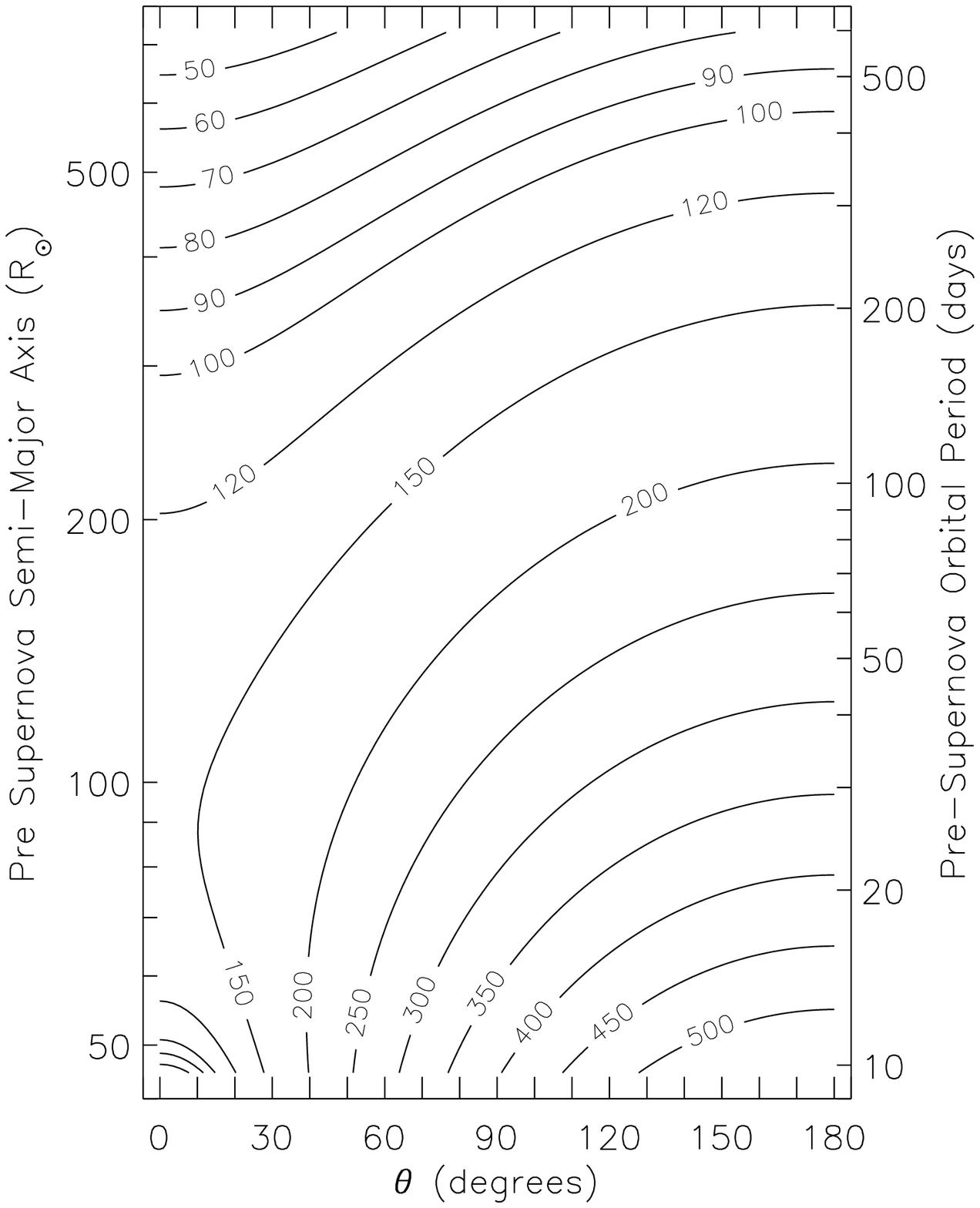,width=3.1in}\caption{Kick velocity (km$~{\rm s}^{-1}$) 
required to produce the present orbit for
pre-supernova systems with a companion mass of $11.3 M_\odot$, and a neutron
star mass of $1.4 M_\odot$. $\theta$ is the angle between the orbital angular
momentum and the Be star spin angular momentum after the supernova explosion.
\label{kicks}}
\end{figure}

Fig. \ref{kicks} shows the kick velocity required for various possible
progenitor systems to have produced the 2S 1845-024 system (see e.g. Kalogera 
et al. (1996\markcite{Kalogera96}) for the relevant equations). Before the
supernova the orbital angular momentum and the companion's spin angular
momentum are aligned, while afterward they are separated by the angle $\theta$.
The calculation assumes a neutron star mass of $1.4M_\odot$, a pre-supernova 
helium star mass of 3.0$M_\odot$, and a companion star mass of $11.3M_\odot$
($i = 60^\circ$). The current eccentricity, period, and semi-major axis are
assumed  unchanged since the supernova. These elements are unaffected by the
tidal interactions and spin-orbital coupling which lead to periastron advance
and precession of the orbital plane and the Be star spin axis (\cite{Lai95}). 
For these parameters a minimum kick of 41 km s$^{-1}$ is required, which occurs
in a 595 day period system with kick directed along the direction of motion. On
the other hand, a kick of 550  km s$^{-1}$ directed opposite to the direction
of motion in a 9.7 day period system could have produced the system. In this
case the Be star spin  and orbital angular momentum are in opposite directions.
The actual origin of the current system probably lies between these extremes.
In particular we wish to emphasis that if typical kick velocities are 
$\sim 250$ km~s$^{-1}$ (\cite{Hansen97}), 
then there is no reason to suppose that
the orbital plane and the Be star equator are closely aligned, as is normally
assumed in the direct wind accretion outburst model (\cite{Waters89}).

\begin{figure}[t]
\psfig{file=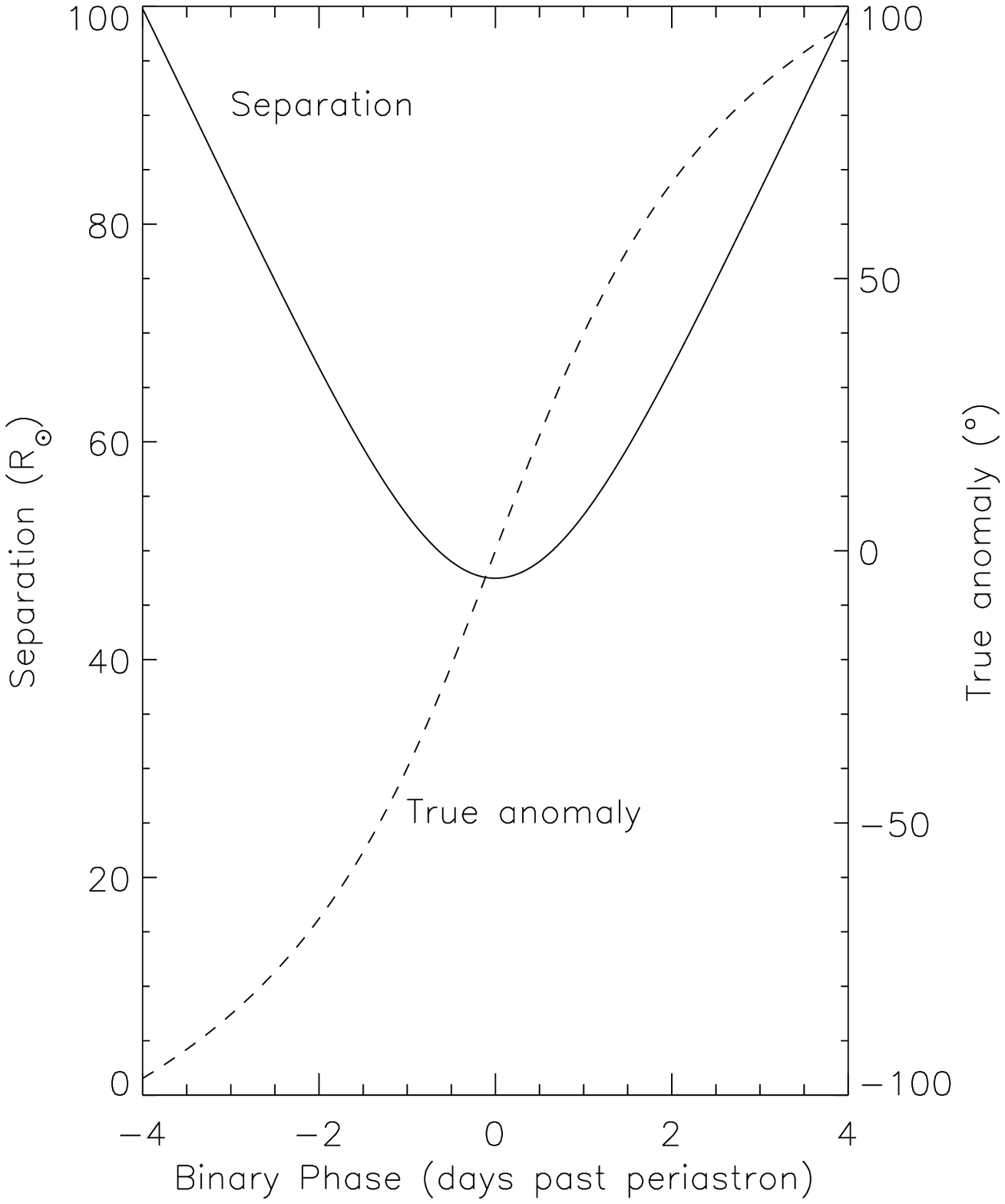,width=3.1in}
\caption{The neutron star's separation from its companion
and the true anomaly, for times near periastron passage. The separation is
calculated assuming a neutron star mass of 1.4$M_\odot$, and an inclination
of $60^\circ$.\label{true_anomaly}}
\end{figure}

We propose that the pulsar's orbit is inclined to its companion's equator, and
that the narrow feature in the outburst profile (Fig. \ref{avgoutburst}) 
near the onset of the outburst is due to direct wind accretion during the
pulsar's passage through the companion's circumstellar envelope.  Fig.
\ref{true_anomaly} shows the pulsars separation from its companion, and its
true anomaly, for times close to periastron. The geometry is changing rapidly,
with the separation changing by a factor of two in less than four days, and 
the true anomaly changing by $43^\circ$ per day at periastron.  If the angle
between periastron and the node of the orbit with the companion's equator is
less than $40^\circ$, then passage through the equatorial plane will occur
within a day of periastron passage.  Be star circumstellar envelopes have
half-angles of $<~10^\circ$ (see \cite{Hanuschik96,Quirrenbach97,Wood97}).  The
passage through the circumstellar envelope would therefore take less than a day
for orbits inclined from the companion's equator by more than $15^\circ$.
Direct wind accretion during the passage could result in the onset feature in
the profile. At this time the accretion disk could also be forming.
Alternatively a disk could already exist (\cite{Bildsten97}), with enhanced
accretion initiated near periastron due to tidal torques and newly added 
material.

This neutron star's wide and eccentric orbit about a massive star is similar to
two radio pulsars PSR B1259-63 (\cite{Johnston92}) and PSR J0045-7319 
(\cite{Kaspi94}). Both of these pulsars are in highly eccentric orbits
($e=0.8698, 0.8080$) about $8-10 M_\odot$ main sequence stars with periods of
1236 and 51.2 days respectively.  Bildsten et al. (1997) had earlier noted the
lack of accreting pulsars with similarly large eccentricities, and conjectured
it was a selection effect due to the difficulties of repeatedly observing long
period systems. 2S 1845-024 begins to fill in this gap. As we have suggested
for 2S 1845-024, the orbital plane of PSR J0045-7319 is believed to be
inclinated from its companion's equator (\cite{Lai95,Kaspi96}).  However, given
the its low spin frequency, it is unlikely that 2S 1845-024 would
be a radio pulsar now or in the future.  

M.H.F. acknowledges support from NASA grant NAG 5-4238. L.B. and T.A.P. 
acknowledge support from NASA grant NAGW-4517. We thank the anonymous referee
for helpful comments, which have improved the text.

\end{document}